# The Near-Sun Dust Environment: Initial Observations from Parker Solar Probe

J. R. Szalay[1], P. Pokorný[2,3], S. D. Bale[4,5], E. R. Christian[2], K. Goetz[6], K. Goodrich[4,5], M. E. Hill[7], M. Kuchner[2], R. Larsen[8], D. Malaspina[9], D. J. McComas[1], D. Mitchell[7], B. Page[4,5], N. Schwadron[10]

[1]Department of Astrophysical Sciences, Princeton University, Princeton, New Jersey, USA
[2]NASA Goddard Spaceflight Center, Greenbelt, MD, 20771, USA
[3]The Catholic University of America, Department of Physics, Washington, DC 20064
[4]Space Sciences Laboratory, University of California, Berkeley, CA, USA
[5]Physics Department, University of California, Berkeley, CA, USA
[6]School of Physics and Astronomy, University of Minnesota, Minneapolis, MN, USA
[7]Johns Hopkins University Applied Physics Laboratory, Laurel, MD 20723, USA
8 Red Rocks Community College, Lakewood, CO, USA
[9]Laboratory for Atmospheric and Space Physics, University of Colorado, Boulder, CO, USA
[10]University of New Hampshire, Durham, NH, 03824, USA

## Abstract

The Parker Solar Probe (PSP) spacecraft has flown into the most dense and previously unexplored region of our solar system's zodiacal cloud. While PSP does not have a dedicated dust detector, multiple instruments onboard are sensitive to the effects of meteoroid bombardment. Here, we discuss measurements taken during PSP's first two orbits and compare them to models of the zodiacal cloud's dust distribution. Comparing the radial impact rate trends and the timing and location of a dust impact to an energetic particle detector, we find the impactor population to be consistent with dust grains on hyperbolic orbits escaping the solar system. Assuming PSP's impact environment is dominated by hyperbolic impactors, the total quantity of dust ejected from our solar system is estimated to be $1-14$ tons/s. We expect PSP will encounter an increasingly more intense impactor environment as its perihelion distance and semi-major axis are decreased.

## 1. Introduction

The Parker Solar Probe (PSP) mission [Fox et al. 2014] explores the inner-most region of our solar system's zodiacal dust disk. While PSP does not carry a dedicated dust detector, its various instruments can each make direct and/or indirect measurements of the local and near-Sun dust environment. It carries four instrument suites: FIELDS [Bale et al. 2016], making measurements of the local electric and magnetic fields; ISꓳIS [McComas et al. 2016], an energetic particle detector suite; SWEAP [Kasper et al. 2016], a solar wind instrument; and WISPR [Vourlidas et al. 2014], a wide-field imager.

The dust distribution in the inner solar system is continually evolving due to various sources and sinks of material. Material shed from comets and asteroids throughout the solar system slowly spirals toward the Sun under the influence of Poynting-Robertson and solar wind drag [Burns et al. 1979]. These dust grains form a population of bound orbits [e.g. Nesvorný et al. 2010; Nesvorný et al. 2011a, 2011b; Pokorný et al. 2014] whose orbits circularize during their transit towards the Sun. Mutual collisions serve to fragment dust grains into smaller fragment particles, which more readily feel the outward force of solar radiation pressure. These smaller fragments are often expelled from the solar system in unbound orbits as β-meteoroids [e.g. Berg and Grün, 1973, Zook and Berg, 1975; Wehry and Mann, 1999] and serve as the most efficient loss mechanism for the zodiacal cloud [Grün et al. 1985a; 1985b]. The relative ratio of bound and β-meteoroids remains an open issue in understanding the evolution of the zodiacal cloud [Mann et al. 2004]. Additionally, a population of very small nanometer sized dust grains can be picked up and entrained in the solar wind [e.g. Czechowski and Mann, 2010; Juhasz and Horányi, 2013; O'Brien et al. 2014, 2018], mass-loading the solar wind.

There have been a number of direct and indirect dust measurements in the inner solar system. In-situ dust measurements from Pioneer 8 & 9 [e.g. Berg and Grün, 1973; Grün et al. 1973], HEOS-2 [e.g. Hoffmann et al. 1975a,b], Helios [e.g. Leinert et al. 1978, 1981; Grün et al. 1980; Altobelli et al. 2006], and Ulysses [e.g. Wehry and Mann 1999; Landgraf et al. 2003 Sterken et al. 2015; Strub et al. 2019] have found three separate populations of inner solar system dust: 1) larger dust grains on bound orbits spiraling into the Sun under of influence of Poynting-Robertson drag, 2) smaller dust grains on hyperbolic orbits escaping the solar system (β-meteoroids), and 3) interstellar dust grains transiting our solar system. Before PSP, Helios probed the dust environment nearest to the Sun between 0.3 – 1.0 au. Remote sensing measurements of the zodiacal light from Helios indicates number density varies as $r^{-1.3\pm0.2}$ [Leinert et al. 1978, 1981]. Since PSP does not have a dedicated dust detector, its primary in-situ method of impact detection is by identifying voltage spikes in the FIELDS data products. There is a large heritage of dust detections with antenna from a number of missions: Voyager 2 [Gurnett et al. 1983], Vega [Laakso et al. 1989], DS-1 [Tsurutani et al. 2004], Wind [Malaspina et al. 2014; Malaspina and Wilson, 2016; Kellogg et al. 2016], MAVEN [Andersson et al. 2015], and STEREO [Malaspina et al. 2015].

In this study, we focus on observations from the FIELDS and ISΘIS instruments. We note that the effect of mass loading due to nanodust picked up and entrained in the solar wind may be detectable by the SWEAP instrument [e.g. Rasca et al. 2014a, 2014b], but will not be discussed further here. We also do not address interstellar dust grains, as the flux of these particles in the inner solar system is expected to be significantly reduced due to radiation pressure [e.g. Sterken et al. 2015]. Additionally, WISPR can make both remote sensing measurements of the more global dust distribution at a range of observable heliocentric distances as well as directly imaging the impact ejecta products from impacts to the PSP spacecraft body. Here, we describe initial observations of direct dust impacts with PSP. In Section 2, we describe the populations of dust impacting the PSP spacecraft and their specific discriminating characteristics. We then compare impactor models to the PSP observations in Section 3. In Section 4, we conclude with a discussion on the implication of these measurements with respect to the zodiacal cloud and highlight how future measurements with PSP may further shed light on the evolution of the zodiacal dust distribution.

## 2. Impact Flux Model

In this section, we describe the model used to compare meteoroid fluxes with PSP measurements. In Section 2.1 we begin by describing the dynamics of the two populations of dust grains considered to illustrate the changing impactor environment encountered by PSP throughout its orbit. We describe our three-dimensional, dynamical model in Section 2.2 and discuss its results in Section 2.3.

### 2.1 Impact Environment for Circular and β-meteoroids

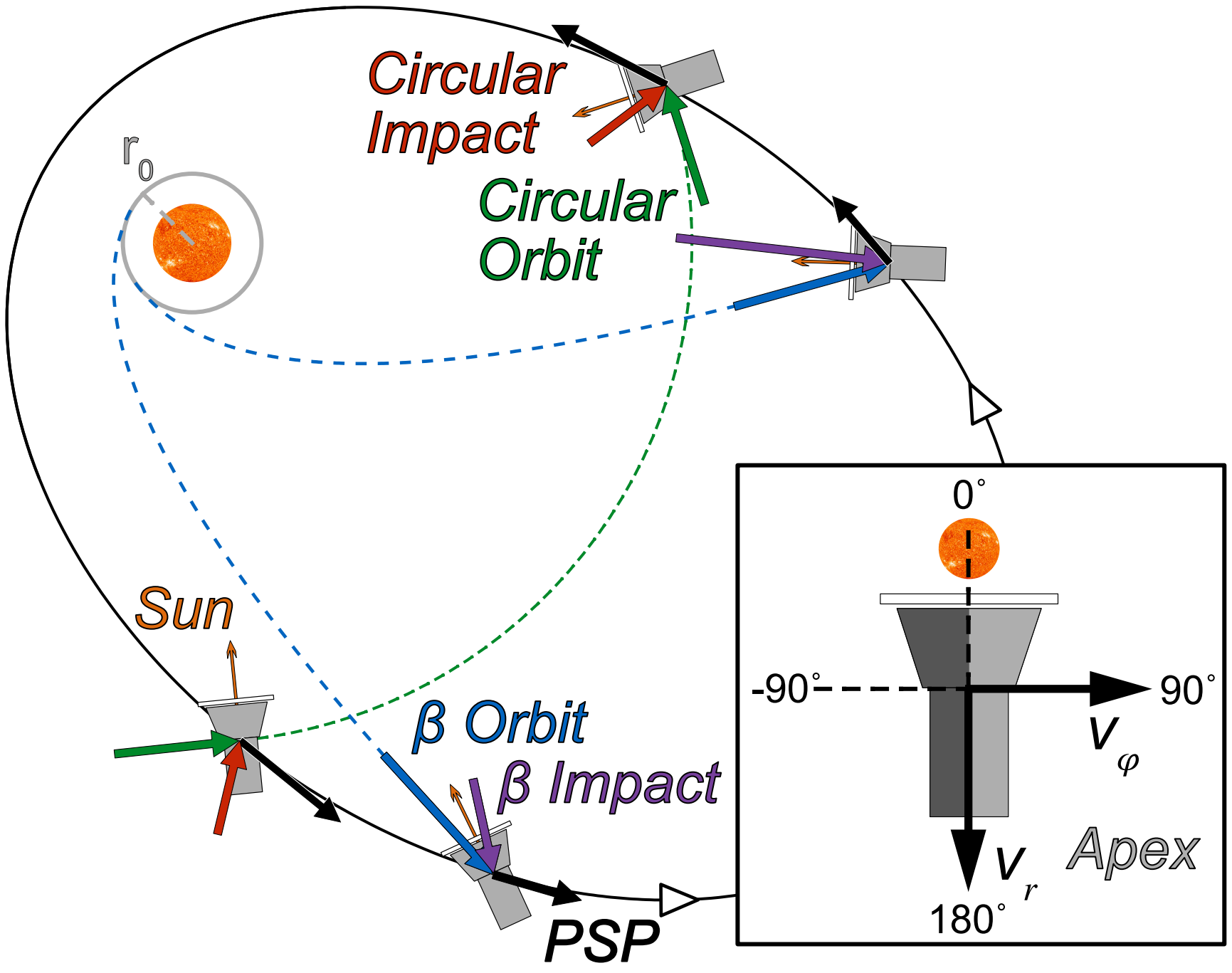

**Figure 1.** Schematic showing vectors for the PSP velocity (black), circular dust velocity vector (green), circular impact velocity (red), β velocity (blue), and β impact velocity (purple) for Orbit 2. The direction to the Sun is shown with the small orange arrows. The inset in the bottom right shows the angle conventions used in this work, in the $r, \varphi$ coordinate system.

Figure 1 outlines the impact geometry for the two populations considered in this study. PSP's eccentric orbit (orbit 2 shown here) causes it to experience a varying impact environment throughout its transit around the Sun. The PSP velocity vector is denoted with black arrows over the small PSP representations. Velocity vectors are shown for circular, bound orbits in green and β-meteoroids in blue. Both types have example trajectories shown with dashed lines. The apparent impact vector that PSP will experience from each source is the vector difference of the dust and PSP velocities, shown for both circular (red) and β-meteoroids (purple). The inset defines the coordinate system used throughout this study, where the apex hemisphere (the hemisphere containing the azimuthal velocity component) corresponds to angles of 0˚-180˚, with the direction pointing from PSP to the Sun fixed at 0˚.

Under certain assumptions described below, impact speeds and directions between the spacecraft and the dust populations can be analytically determined from their velocity vectors. From conservation of energy, the magnitude of an orbiting objects' velocity (the vis-viva equation) is,

$$v = \sqrt{\mu\left(\frac{2}{r} - \frac{1}{a}\right)}, \qquad (1)$$

where $v$ is heliocentric speed, $\mu = GM = 1.38 \times 10^{20}$, $r$ is the heliocentric radial distance, and $a$ is the semi-major axis. Assuming no mass is lost during the object's motion, from conservation of specific angular momentum $h = rv_\varphi = \sqrt{\mu a(1 - e^2)}$, hence the azimuthal ($\varphi$) speed is

$$v_\varphi = \frac{\sqrt{\mu a(1 - e^2)}}{r}. \qquad (2)$$

For motion in the orbital plane, the radial speed is then,

$$v_r = \pm\sqrt{v^2 - v_\varphi^2}, \qquad (3)$$

which has negative values on the inbound arc and positive values on the outbound arc. A dust grain on a perfectly circular, bound, prograde orbit with $r = a$ and $e = 0$ has velocity components $v_{c\varphi} = \sqrt{\mu/r}$ and $v_{cr} = 0$, listed in Table 1. Similarly, the spacecraft's radial and azimuthal speeds may be calculated by Eqs. 1-3, also given in Table 1. Note, we have assumed circular orbiting dust grains are on prograde orbits, with positive azimuthal velocity. While there also exists a population of dust grains on retrograde orbits, their number densities are significantly reduced due to their high collision probability with the dense zodiacal cloud in the inner solar system [e.g. Steel and Elford, 1986]. Additionally, modeling efforts predict the population of dust grains on retrograde orbits comprise <10% of the zodiacal cloud [e.g. Nesvorny et al. 2010; 2011a; 2011b]. Given their difficultly to constrain with current measurements and their expected low contribution to the total zodiacal cloud, we do not consider the population of dust grains on retrograde orbits in this study.

For β-meteoroids, we follow the analysis and assumptions outlined in Zook and Berg [1975]. To briefly summarize, we assume β-meteoroids are generated at an initial distance $r_0$, most likely due to catastrophic collisions of larger dust grains, which we further assume are initially on perfectly circular, prograde orbits. We use a value of $r_0 = 5R_\odot$ as a canonical source location, guided by existing dust measurements and modeling efforts [e.g. Mann et al. 2004], and discuss the sensitivity of our results to this assumption in the discussion. The collision products are expected to initially retain the same kinetic energy and orbital angular momentum as their parent products. Additionally, unlike the orbits for the circular assumption, all β-meteoroids with $\beta > 0.5$ have positive orbital energy, such that they travel on unbound, hyperbolic trajectories. Here, β is the ratio of the radiation pressure and gravitational forces from the Sun [e.g. Zook and Berg, 1975; Burns et al. 1979]. Throughout this work, we investigate the dynamics of β-meteoroids with $\beta > 0.5 - 1.2$, which corresponds to a dust grain radius of 0.6 μm down to 0.2 μm for a bulk particle density of 2 g/cc [Zook and Berg, 1975] and covers a reasonable size range below the smallest particles expected to be in bound orbits within the zodiacal cloud.

The initial specific angular momentum is given by $h_0 = r_0 v_{\beta\varphi_0} = r_0\sqrt{\mu/r_0} = \sqrt{\mu r_0}$. From conservation of specific angular momentum, $h = r v_{\beta\varphi} = \sqrt{\mu r_0}$, the angular speed of β-meteoroids at any distance is

$$v_{\beta\varphi} = \frac{\sqrt{\mu r_0}}{r}. \qquad (4)$$

Conserving kinetic energy before and after the β-meteoroid is generated gives a relation for the magnitude of the velocity (see Eq. 23 and surrounding text of Zook and Berg [1975]),

$$v_\beta = \sqrt{2\mu\left[\frac{\beta - 1/2}{r_0} + \frac{1-\beta}{r}\right]}, \qquad (5)$$

which can then be combined with Eq. 3 to determine the radial speed at any location,

$$v_{\beta \mathrm{r}} = \sqrt{2\mu\left[\frac{\beta - 1/2}{r_0} + \frac{1-\beta}{r} - \frac{r_0}{2r^2}\right]}. \qquad (6)$$

Table 1 summarizes the heliocentric radial and azimuthal speeds for circular and β-meteoroids as well as the spacecraft.

| | $\boldsymbol{v_r}$ | $\boldsymbol{v_\varphi}$ |
|---|---|---|
| **Circular, Bound** | 0 | $\sqrt{\frac{\mu}{r}}$ |
| **β-meteoroid** | $\sqrt{2\mu\left[\frac{\beta-1/2}{r_0}+\frac{1-\beta}{r}-\frac{r_0}{2r^2}\right]}$ | $\frac{\sqrt{\mu r_0}}{r}$ |
| **Spacecraft**<br>– ⟶ Inbound<br>+ ⟶ Outbound | $\mp\sqrt{\mu\left(\frac{2}{r}-\frac{1}{a}-\frac{a(1-e^2)}{r^2}\right)}$ | $\frac{\sqrt{\mu a(1-e^2)}}{r}$ |
| **Circular Impact**<br>+ ⟶ Inbound<br>– ⟶ Outbound | $\pm\sqrt{\mu\left(\frac{2}{r}-\frac{1}{a}-\frac{a(1-e^2)}{r^2}\right)}$ | $\frac{\sqrt{\mu}}{r}\left(\sqrt{r}-\sqrt{a(1-e^2)}\right)$ |
| **β Impact**<br>+ ⟶ Inbound<br>– ⟶ Outbound | $\sqrt{2\mu\left[\frac{\beta-1/2}{r_0}+\frac{1-\beta}{r}-\frac{r_0}{2r^2}\right]}\pm\sqrt{\mu\left(\frac{2}{r}-\frac{1}{a}-\frac{a(1-e^2)}{r^2}\right)}$ | $\frac{\sqrt{\mu}}{r}\left(\sqrt{r_0}-\sqrt{a(1-e^2)}\right)$ |

**Table 1.** Radial and azimuthal speeds for perfectly circular orbiting dust grains, β-meteoroids, and an orbiting spacecraft all in the same plane. The impact vectors $\vec{v}_{imp}=\overrightarrow{v_d}-\overrightarrow{v_{sc}}$ are given in the last two rows.

From these relations, the impact velocity vector can be calculated by transforming the dust velocity vectors to the spacecraft frame, $\vec{v}_{imp} = \overrightarrow{v_d} - \overrightarrow{v_{sc}}$, where $\overrightarrow{v_d}$ is the dust population velocity vector and $\overrightarrow{v_{sc}}$ is the spacecraft velocity vector, given in Table 1. In all these calculations, we assume the spacecraft is exactly orbiting in the same plane as the dust grain orbits. This is reasonably justified as PSP's inclination in small, ~4˚. Hence, the velocity vectors can be fully described with the radial and azimuthal components.

In addition to the impact speeds, the impact angle can be calculated as

$$\theta = \tan^{-1}\left(\frac{v_{imp,\varphi}}{v_{imp,r}}\right), \qquad (7)$$

where $\theta = 0°$ corresponds to an impact coming from the solar direction outward following the convention in Figure 1. Figure 2 shows the impact speeds and directions for both populations throughout all PSP orbit families.

For circular impactors, the impact vector will rotate completely throughout 360˚ over a full orbit. There are two key locations along the orbit where the impact velocity is purely radial and has no azimuthal component, such that $v_{imp,\varphi} = 0$. This is where the impact vector from perfectly circular orbiting dust grains transitions between ram and anti-ram hemispheres on the spacecraft. It occurs at a distance of

$$r_k = a(1 - e^2), \qquad (8)$$

which we term the "Kepler distance", where the orbital elements *a* and *e* are those of the spacecraft. Intuitively, this is the location where the spacecraft intersects impacting particles with the same specific angular momentum. This distance is also where the spacecraft experiences the peak impact speed for circular dust grains,

$$v_{imp,k} = \sqrt{\mu\left(\frac{1}{r_k} - \frac{1}{a}\right)} = \sqrt{\frac{\mu}{a}\left(\frac{e^2}{1 - e^2}\right)}, \qquad (9)$$

and is therefore an important location for hazard estimates. Outside the Kepler distance, impacting dust grains have larger specific angular momentum and are "catching up" to the spacecraft, hitting it from the anti-apex hemisphere. Inside this distance, the opposite occurs, where the spacecraft has larger specific angular momentum than the dust grains on circular orbits impacting it and therefore experiences impacts from the apex hemisphere. Table 2 lists the Kepler distance and impact speed for all currently projected PSP orbits, along with relevant PSP orbital parameters. The PSP orbits are grouped by their common orbital characteristics calculated at each perihelion as in Figure 2.

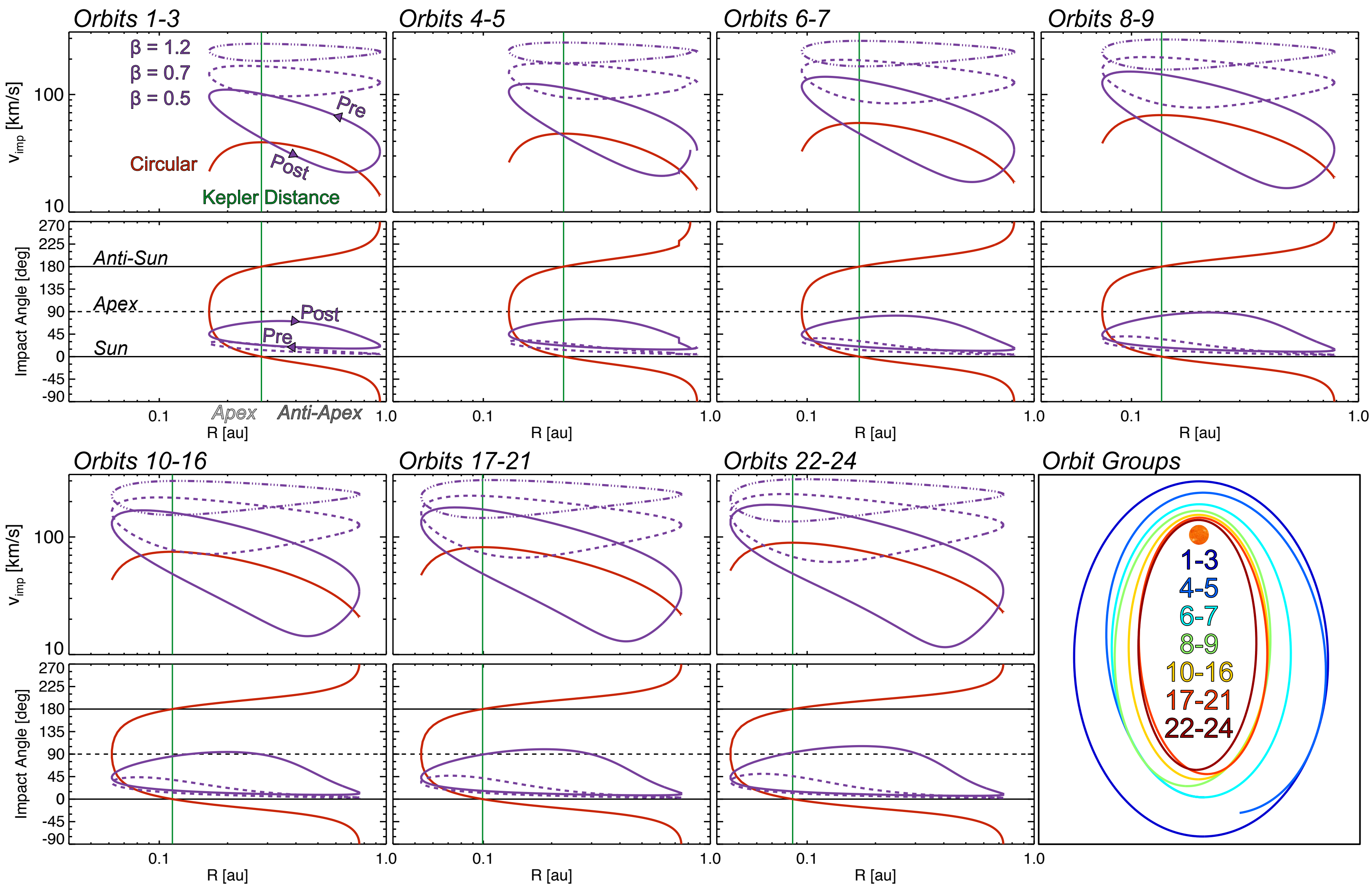

**Figure 2.** Impact speed and direction for circular (red) and beta-meteoroids (purple) for all orbit families.

Helios measurements indicated the zodiacal cloud scales as $n_c(r) \propto r^{-1.3}$ [Leinert et al, 1978], which is considered to be predominantly composed of bound dust grains orbiting in approximately circular orbits [e.g. Aimanov et al. 1995]. This radial trend was also reproduced with a dynamical meteoroid model of dust grains shed from Jupiter Family Comets [Pokorny and Kuchner, 2019]. However, β-meteoroids are most likely the main sink for material to be lost from the zodiacal cloud [Grün et al. 1985a] and the relative contributions of bound vs β-meteoroids is still an open issue in understanding the evolution of the zodiacal cloud [Mann et al. 2004]. In order to compare trends between bound and β-meteoroids, we can estimate the expected radial scaling for the population of β-meteoroids discussed in this work. Conservation of mass dictates $n v_r r^2 = const.$ We focus on heliocentric distances larger than $r_0$, outside the source region for β-meteoroids and assume all β-meteoroid production occurs inside this distance. Substituting for the radial speed from Table 1, the density is then

$$n_\beta(r) \propto \frac{1}{v_r r^2} \propto \frac{1}{r^2}\left[1 + \left(\frac{r_0}{r}\right)\left(\frac{1-\beta}{\beta - 1/2}\right)\right]^{-1}. \qquad (10)$$

This reduces to approximately $n_\beta(r) \propto r^{-2}$ when

$$\left(\frac{r_0}{r}\right)\left(\frac{1-\beta}{\beta - 1/2}\right) \lesssim 1. \qquad (11)$$

Solving for when the left-hand side of Eq. 11 equals unity gives the minimum value for β to satisfy this relation,

$$\beta = \frac{r_0/r + 1/2}{1 + r_0/r}. \qquad (12)$$

When $r = r_0$, this relation is strictly satisfied for $\beta \geq 0.75$. Outside 0.1 au, values of $\beta \gtrsim 0.6$ for $r_0 = 5R_\odot$ also satisfy this relation. Hence, at $r \geq 0.1$ au, the population of β-meteoroids for $\beta \gtrsim 0.6$ has a number density that call be well approximated to scale like $n_\beta(r) \propto r^{-2}$. This is in contrast to the bound dust grain density scaling of $n_c(r) \propto r^{-1.3}$, which is due to spatially distributed source region and collisional effects for this population of orbiting dust grains leading to a shallower exponent [Leinert et al. 1983].

| **Orbit** | **a** | **e** | **i** | **$r_{ph}$** | **$v_{ph}$** | **$r_k$** | **$v_{imp,k}$** |
|---|---|---|---|---|---|---|---|
| | au | | deg | au ($R_\odot$) | km/s | au ($R_\odot$) | km/s |
| **1-3** | 0.5520 | 0.6994 | 3.361 | 0.16 (36) | 95 | 0.28 (60) | 39 |
| **4-5** | 0.5021 | 0.7418 | 3.395 | 0.13 (28) | 109 | 0.23 (49) | 46 |
| **6-7** | 0.4557 | 0.7922 | 3.384 | 0.095 (20) | 130 | 0.17 (36) | 57 |
| **8-9** | 0.4284 | 0.8266 | 3.395 | 0.074 (17) | 150 | 0.14 (29) | 66 |
| **10-16** | 0.4112 | 0.8498 | 3.388 | 0.062 (13) | 163 | 0.11 (25) | 74 |
| **17-21** | 0.3990 | 0.8667 | 3.394 | 0.053 (11) | 176 | 0.099 (21) | 81 |
| **22-24** | 0.3885 | 0.8819 | 3.394 | 0.046 (9.9) | 191 | 0.086 (19) | 89 |

**Table 2.** PSP Orbit characteristics.

### 2.2 Dynamical Model for Bound Dust Grains

Having described the orbital dynamics and impact geometry for idealized orbits, we employ a three-dimensional, dynamical model to more comprehensively predict the dust environment PSP encounters. This model follows the individual trajectories of a large number of particles shed from four separate dust sources, described in Pokorny & Kuchner [2019]: Jupiter Family Comets (JFC), Halley Type Comets (HTC), Oort Cloud Comets (OCC), and Main Belt Asteroids (MBA). All model populations include dust grains in the size range of 0.3 μm to 1 mm in radius.

The dynamical model is constrained with a large number of observations. The total mass flux is normalized to be consistent with the terrestrial mass flux of 43 tons/day at 1 au [Carrillo-Sánchez et al. 2016]. The orbital element distribution well-matches that of meteors observed at Earth from CMOR [Campbell-Brown, 2008] and SAAMER [Janches et al., 2015] and the spatial number density is consistent with Helios observations of the zodiacal cloud's radial density distribution [Leinert et al. 1978]. For the size-frequency distribution, we use previously determined values for the meteoroid cloud at 1 au [Love et al. 1993; Cremonese et al. 2012]. Finally, the latitudinal profile is found to match observations of zodiacal cloud from a variety of remote measurements [e.g. Dermott and Nicholson, 1989; Kelsall et al. 1998; Rowan-Robinson and May, 2013]

This model, along with its previous iterations, has been successfully employed to understand spacecraft measurements throughout the inner solar system. At 1 au, it has been used to help understand and extend lunar impact ejecta measurements from the LADEE mission [Janches et al. 2018; Pokorný et al. 2019] as well as to compare the meteoroid impact flux and directions on the LISA Pathfinder spacecraft [Thorpe et al. 2019]. At 0.7 au, it was used to explain the presence of a dust ring in Venus' orbit and made predictions on the quantity dust shed from yet-undiscovered asteroids co-orbiting with Venus [Pokorny & Kuchner 2019]. At 0.3 au, it was used to interpret the asymmetries in Mercury's exospheric densities observed by the MESSENGER spacecraft [Pokorný et al. 2017, 2018]. Hence, it is a robust model in describing many dust related phenomena throughout the solar system, so far from 0.3 - 1.0 au, and we employ it to better understand dust related PSP measurements.

### 2.3 Dynamical Model results

By integrating a large number of trajectories, the model can generate flux maps given any position and velocity vector in the solar system. Figure 3 shows an example of such a flux map, which is displayed in an outward looking Mollweide projection of the dust number flux for the entire sky at PSP and follows the angle conventions in Figure 1. Each line indicates 30˚ latitude/longitude. The equator is aligned with PSP's orbital plane and it is centered on the solar direction. The example shows the impactor flux when PSP is at perihelion, therefore the peak impactor flux is coaligned with the ram (PSP velocity) direction at +90˚ from the solar direction.

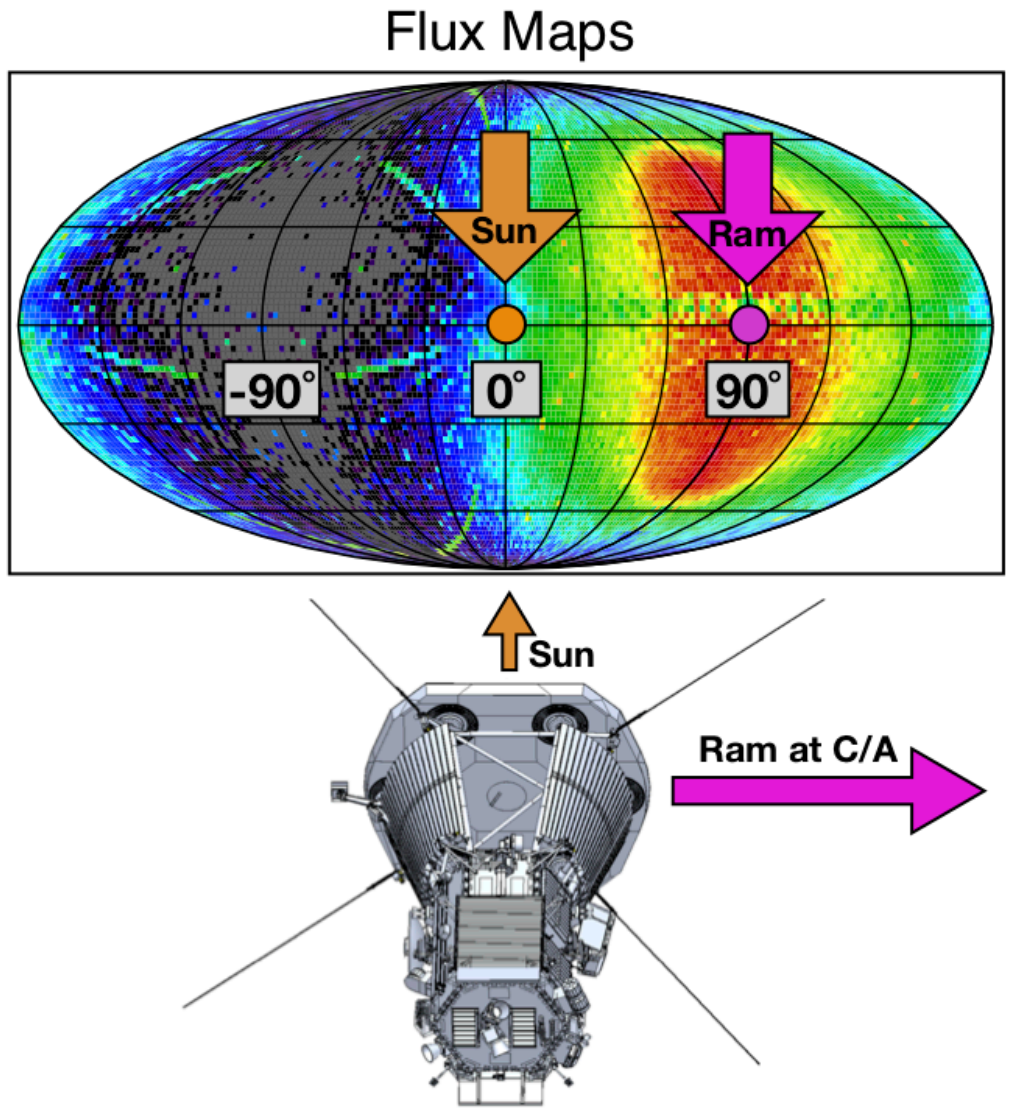

**Figure 3.** Mollweide projection for the impact predictions. The frame is centered on the solar direction with the equatorial plane containing the solar direction and PSP velocity vector.

Dust grains shed from JFCs are expected dominate the number flux inside 1 au [Pokorny & Kuchner 2019]. However, we also modeled the impactor environment from dust grains shed by main belt asteroids. The model predicts asteroidal impactors have fluxes that are at least an order of magnitude lower than those of JFCs and therefore we do not consider their effects further. The same applies for the remaining long-period cometary sources (HTCs and OCCs). Figure 4 shows the JFC model flux maps for the pre- and post-perihelion arcs of PSP's orbit 2 at $r = [0.17, 0.2, 0.3, 0.4, 0.5, 0.6, 0.7, 0.8, 0.9]$ au. The red arrows indicate the circular dust impactor direction with the length proportional to the impact speed. A small simplified version of the PSP spacecraft along with its sun-pointing orientation is shown at each location as well. Small white dots in each map indicate the average impactor flux direction from the dynamical model and the orange circles show the solar direction.

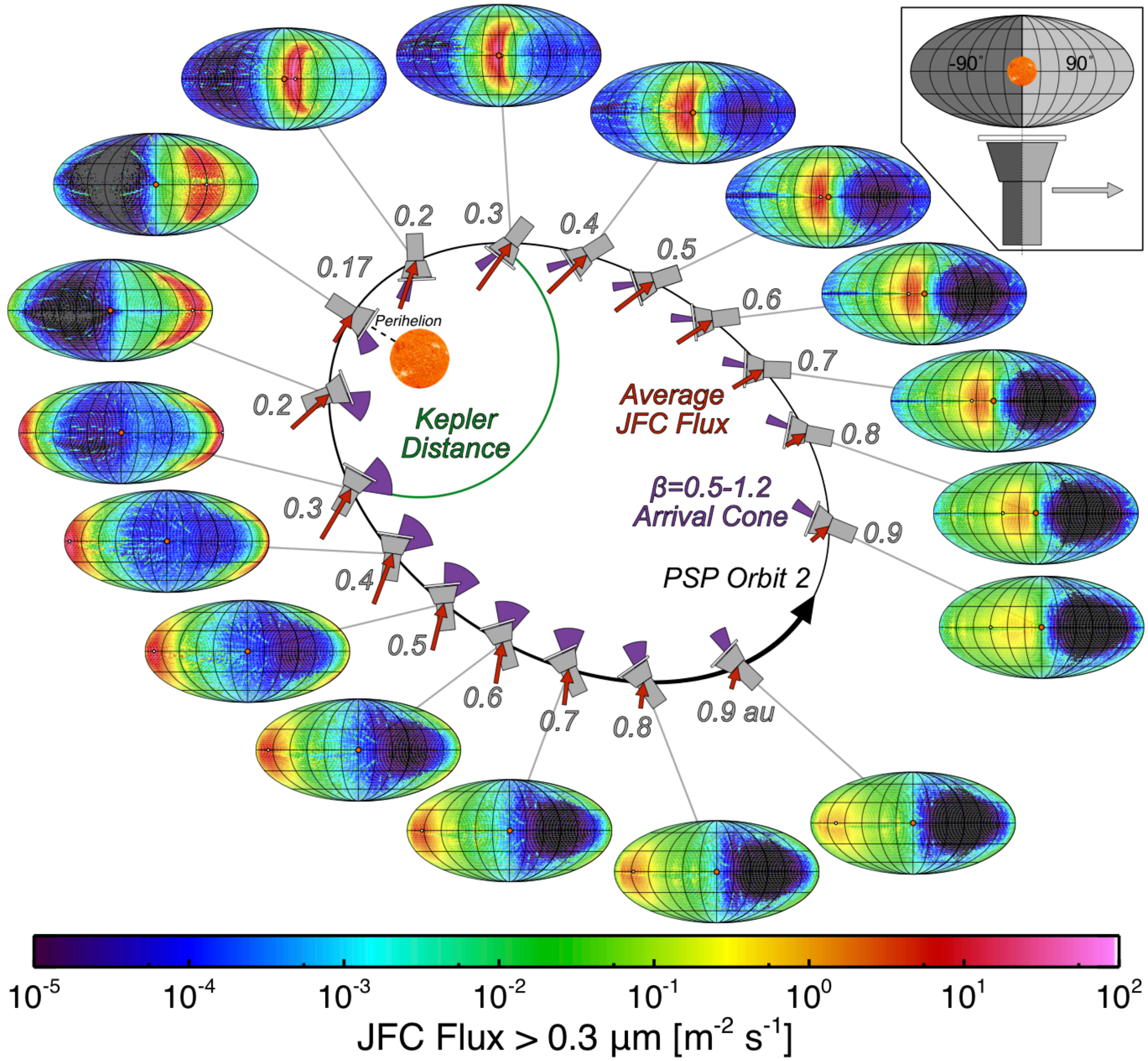

**Figure 4.** Impactor fluxes along PSP's Orbit 2. Red arrows indicate the average dust impact direction for each PSP location and purple wedges show the arrival direction for β-meteoroids with β = 0.5-1.2. A logarithmic color scale is used to show the extent of impactor flux variation.

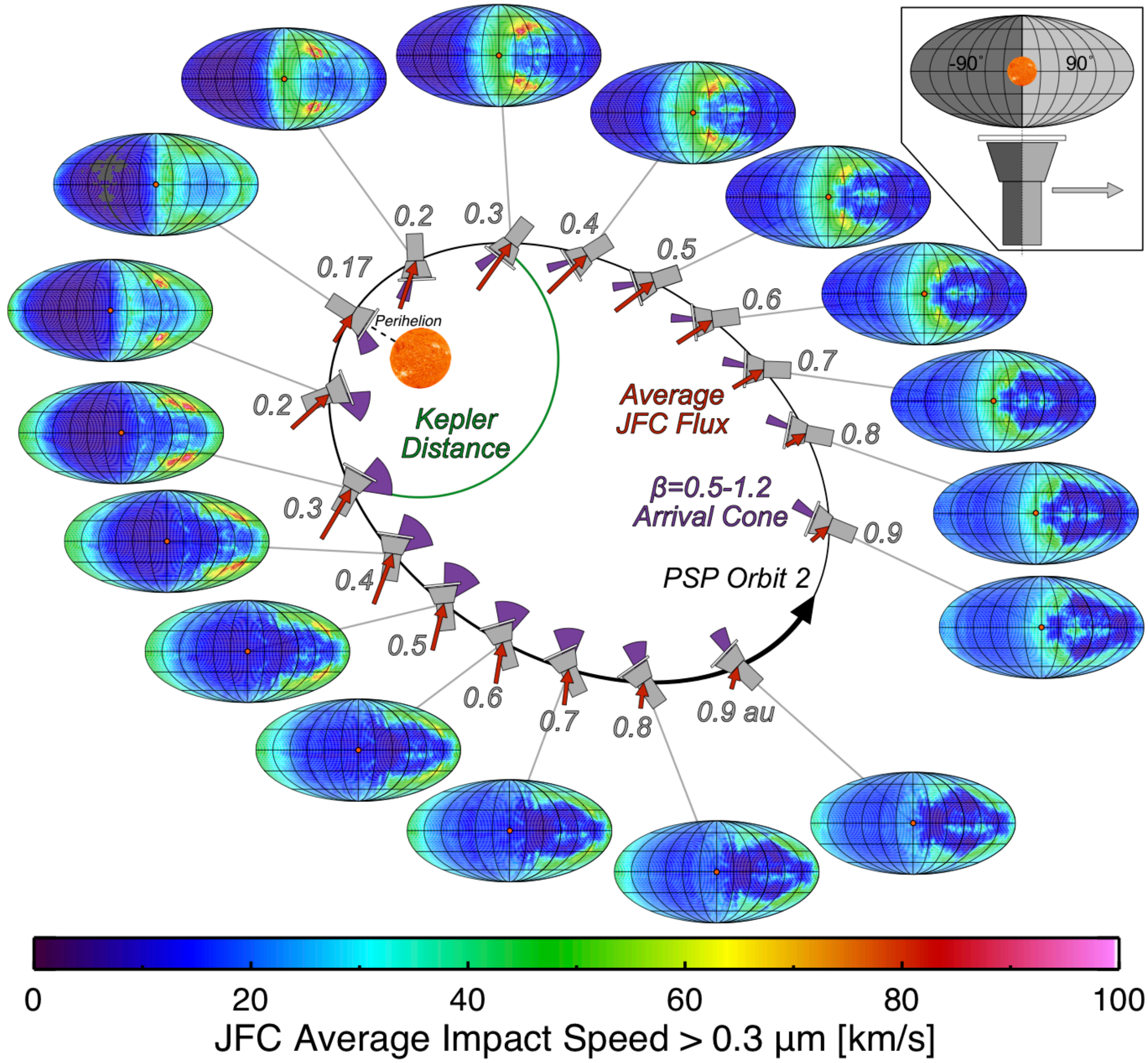

**Figure 5.** JFC average impact speed along PSP's Orbit 2. See description in Figure 4 for additional information.

Figure 5 shows the dynamical model average speed across the entire sky in the same configuration as Figure 4. The overall speed distribution follows similar trends to the simple circular case, where a peak in total impact speed occurs near the Kepler distance before and after perihelion. Both at perihelion and aphelion, there is a local minimum in overall impactor speeds. An additional population of impactors is revealed from the JFC population that is not represented by the circular assumption. Between latitudes of 30˚ - 60˚ there is a significant peak in average impactor speed between longitudinal angles of 30 ˚ - 150˚ on the apex hemisphere. This enhancement is due to the fact that the JFCs and their shed dust grains are distributed up to 30˚ inclined from the ecliptic, leading to much higher impact speeds from the vector velocity addition of such a population. It is generated by the same dynamical process that creates the toroidal sporadic meteoroid source observed from radar and visual observations at Earth [Pokorny et al. 2014]. While diminished in flux, this population has significantly higher average impact speeds, approaching 100 km/s at some portions of the orbit, and could be an important source to consider for hazard estimates to the spacecraft. Additionally, if PSP were able to constrain the relative fluxes of inclined, bound dust grains, it would improve our understanding of how such a toroidal population is generated and evolves.

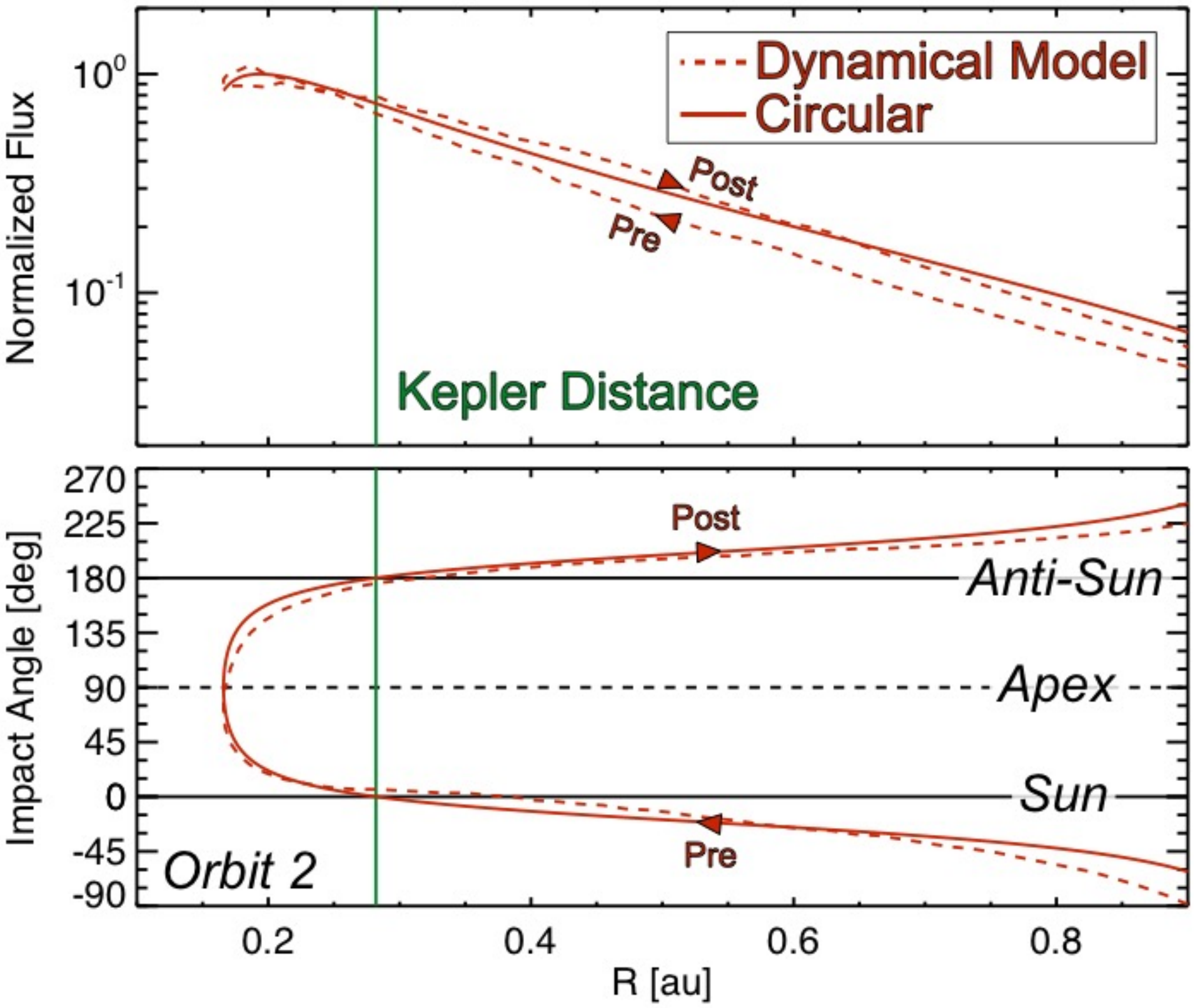

**Figure 6.** Comparison of dynamical model (dashed) and perfectly circular assumption (solid) for the normalized flux (top) and average impact angle (bottom) for PSP orbit 2.

Having implemented the dynamical model, we investigate differences between it and simply assuming all dust grains are on circular orbits with $n(r) \propto r^{-1.3}$. Figure 6 shows normalized impactor fluxes (the product of density and impact speed) and average impact angle to PSP for orbit 2. Overall, the dynamical model and circular assumption produce similar predictions. Both predict a peak in impactor flux occurs before perihelion, due to depression in impact velocity near close approach. Additionally, both predict similar flux slopes as a function of radial distance.

However, unlike the circular case, which assumes all impacting dust grains and PSP are in the same plane, the dynamical model fully accounts for PSP's orbit, where the inclination is now considered. In contrast to the circular case, where the expected impact flux only depends on radial distance (due to conservation of angular momentum), the dynamical model predicts small asymmetries between the inbound/outbound, pre/post-perihelion arcs. Both PSP's small inclination as well as the simulated JFC dust orbital element distribution contribute to these asymmetries. For example, the model predicts a slightly larger flux on the post-perihelion arc. Additionally, it predicts impacts transition between apex and anti-apex hemispheres at slightly larger radial distances than the Kepler distance, at $r = 0.39$ au inbound and $r = 0.31$ au outbound, compared to $r_k = 0.28$ au for this orbit. As such, asymmetries in directionality data for impacts to PSP may encode key information of the orbital distribution of impacting dust grains.

# 3. Comparison to PSP Data

### 3.1 Expected trends

While the lack of a dedicated dust detector onboard PSP limits our ability to characterize the impactor fluxes, there are two key diagnostic measurements that can guide the interpretation of dust-related PSP measurements: the radial trend and average impact direction. As previously discussed, dust grains on nearly circular bound orbits and β-meteoroids have different characteristic expected radial trends and directionalities. Table 3 summarizes these differences, where the radial trend for dust grains on circular orbits is from Leinert et al. [1978].

| | **Radial Trend** | **Impact Direction** |
|---|---|---|
| **Circular** | $n(r) \propto r^{-1.3}$ | Rotates 360° throughout orbit. Transitions between apex/anti-apex near the Kepler distance. |
| **β-meteoroids** | $n(r) \propto r^{-2}$ (for $\beta \gtrsim 0.6$) | Always apex hemisphere, predominantly more from the solar direction. |

**Table 3.** Expected radial and impact direction differences between circular and β-meteoroids.

**3.2 FIELDS Data**

The FIELDS instrument [Bale et al. 2016] aboard PSP has multiple data receivers, each with multiple data products capable of registering dust impacts. Here we examine data from the FIELDS antennas as processed by the Time Domain Sampler (TDS) and Digital Fields Board (DFB) receivers.

Electric field instruments on board spacecraft detect dust via impact ionization. When a dust grain strikes a spacecraft surface at high velocity, the grain and some fraction of the spacecraft material struck are vaporized and ionized. The dynamics of the resulting plasma cloud (expansion, charge separation, and/or recollection) produce disturbances in the electric potential between electric field instrument voltage probes and the spacecraft body. Such disturbances are often impulsive (lasting a few ms) and can reach high amplitudes up to a few V. They typically do not produce measurable signals in magnetic field instruments. Laboratory measurements of hyper-velocity dust impacts on scaled down electric field instrument systems have confirmed this signal generation mechanism [Collette et al. 2013, 2015, 2016; Nouzák et al. 2018].

Dust impacts have been observed by electric field instruments in the vicinity of the gas giant planets [Gurnett et al 1983, 1986, 1987, 1991, 2004; Tsintikidis et al 1995; Kurth et al 2006; Meyer-Vernet et al 2009, Ye et al. 2014], Mars [Andersson et al. 2015], Earth [Vaverka et al. 2018], comets [Gurnett et al 1986; Meyer-Vernet et al 1986; Laakso et al 1989; Neubauer et al 1990; Tsurutani et al 2003], and in the open solar wind [Meyer-Vernet et al 2009b; St. Cyr et al 2009; Zaslavsky et al. 2012, Malaspina et al 2014; Kellogg et al 2016, Malaspina and Wilson 2016].

Dust impact voltage spikes are distinct from natural short-duration impulsive electric field structures such as electron phase space holes or double layers. Important distinctions are: (i) the waiting time between dust impacts is often far longer than the waiting time between natural impulsive structures, (ii) the amplitudes of dust impacts are often (but are not necessarily) much larger than natural impulsive structures, (iii) dust impact observed amplitudes are often systematically higher when measured with a monopole sensor configuration compared to measurement with a well-balanced dipole configuration [Meyer-Vernet et al. 2014], while the amplitude of natural electric field structures does not depend in this way on measurement configuration, and (iv) dust impacts may be associated with observed images of spacecraft material ejected during the impact [St. Cyr et al. 2009].

The two main potential noise sources for this type of dust detection are instrument configuration changes and electrical discharges between differentially charged surfaces of the spacecraft. While discharges on the spacecraft can generate a signal that mimics a dust impact, such discharges typically cascade and are not distributed in time according to a Poisson distribution. In performing a waiting time analysis, we find the delta time distribution between impacts follows an exponential probability distribution, consistent with a Poisson-process as expected for an impacting dust population [Page et al. submitted]. Additionally, we systematically remove any spurious signals by excluding periods around instrument configuration changes from the impact rate calculation. Periods associated with spacecraft thruster operation are also excluded. Finally, we note that FIELDS impact signals are well correlated with directly imaged impact ejecta products by the WISPR imager [Vourlidas et al. 2016], similar to the correlation made between STEREO impacts and subsequently imaged ejecta [St Cyr et al. 2009].

The TDS samples signals from the FIELDS antennas from a few kHz to ~1 MHz. It produces a peak-detection data product where the largest amplitude signal on a given channel is reported every ~52 seconds (outside of 55 solar radii) or every ~5 seconds (inside 55 solar radii). Isolated peak detections with amplitudes many standard deviations above the instrument noise level are interpreted as dust impacts (see Page et al. [this issue], for details).

The DFB samples signals from the same FIELDS antennas, but between DC and 75 kHz [Malaspina et al. 2016]. It uses the difference between successive outputs of a cascading digital filter bank to produce multiple streams of band pass filtered time series data. The peak amplitude in each band pass bin each ~0.87 seconds is reported when PSP is below 55 solar radii on Orbit 2. This AC-coupled bandpass product has 7 pseudo-logarithmically spaced frequency bins spanning a few hundred Hz to 75 kHz. A dust detection in DFB bandpass data occurs when at least 5 of the 7 bandpass bins for a given ~0.87 sec meet the following criteria: the signal in a given bin is 4 times larger than the signal just before it and just after it. The 5/7 bin criteria selects for structures that are narrow in time and therefore broadband in frequency space.

Both the DFB and TDS produce a limited number of high-cadence time-domain antenna signal waveform captures where dust impact waveforms are fully resolved. These waveforms correspond to times when the TDS and DFB peak-detection products indicate dust impacts. The time-resolved waveforms show the characteristic shapes of dust impacts detected via an electric field instrument, consistent with the prior studies listed above [also see Page et al., this issue].

FIELDS's sensitivity to impact detection depends on the total effective exposed area for which impacts can be registered as well as the efficiency of detection for impacts on various different spacecraft materials and their distance from the antennae. To compare expected model impact rates to FIELDS rates, we must make assumptions about the detection efficiency and total effective area. For simplicity, we assume that FIELDS detects any impact on the spacecraft with equal efficiency, and that the effective detection area is the projected area of all spacecraft surfaces for a given dust impact flux direction. We make this simplifying assumption as these dependencies have not been fully characterized and are difficult to calibrate in flight with the current set of observations. Additional details on the TDS data products, FIELDS antenna location and geometry, and dust impact identification are given in a parallel study [Page et al. this issue].

The top panel of Figure 7 shows the total TDS and DFB dust impact rates (> 50 mV) as a function of time during Orbit 2 inside ~0.7 au, where PSP's heat shield was pointed towards the Sun. The TDS rates are calculated across the two of the FIELDS antennae on the heat shield side of the PSP spacecraft (V3 & V4). DFB rates are only available within 0.25 au of perihelion and are consistent with the TDS rates. With the exception of a few outliers, the impact rate monotonically increases as PSP travels towards perihelion, and decreases monotonically as PSP continues on the outbound, post-perihelion arc of its orbit.

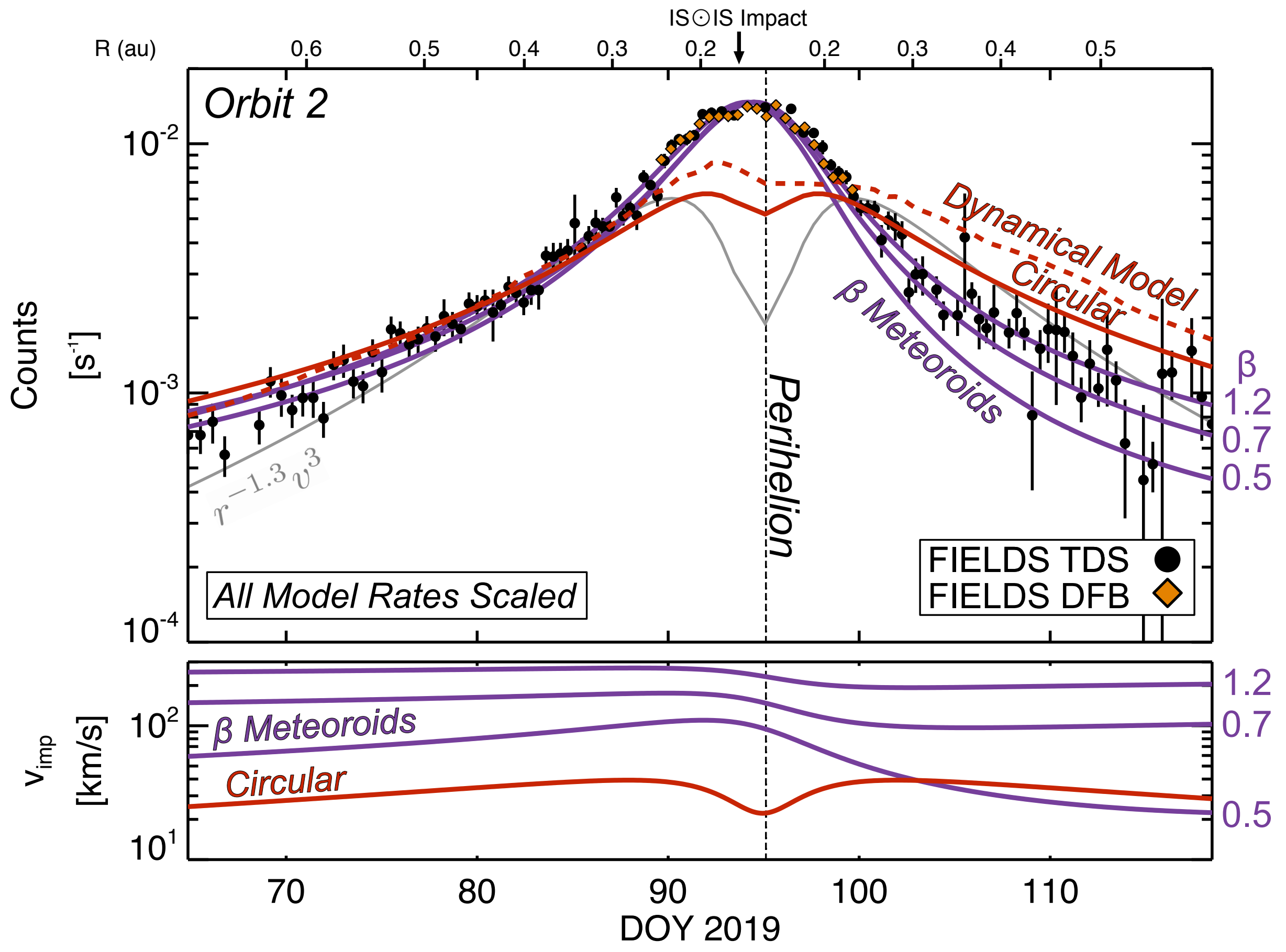

**Figure 7.** Orbit 2 dust summary time series. The top panel shows FIELDS TDS impact rates (black), dynamical model (dashed red), circular model (solid red), β-meteoroids (purple) for β = [0.5,0.7,1.2]. The grey line shows a conservative example of an impact rate for a speed-dependent threshold. The bottom panel shows modeled impact speeds for circular (red) and β-meteoroids (purple). The ISΘIS dust impact is noted with an arrow just before perihelion.

Overlaid on the impact rates are model rates for bound particles from the circular model (red solid) and dynamical model (dashed red) along with three curves for β-meteoroids (purple) for β = [0.5,0.7,1.2]. Model rates are calculated via

$$R = nAv_{imp}, \quad (13)$$

where $n$ is the number density and $A$ is the PSP projected area along the direction of $\overrightarrow{v_{imp}}$, with the exception of the dynamical model as this already gives expected fluxes $F = nv_{imp}$ and only requires multiplication by the effective area $A$. All model rates are scaled to approximately fit the pre-perihelion rates for comparison.

The bottom panel of Figure 7 shows the expected impact speeds for circular and β-meteoroid populations. As previously discussed, the peak impact speed for dust grains on circular orbits occurs at $r_k$, and a local minimum in impact speed exists at perihelion. Since β-meteoroids are all moving on outbound hyperbolic trajectories with $v_{\beta r} > 0$ at all times, the impact speeds follow a different trend. The highest impact speeds for β-meteoroids occur on the pre-perihelion arc where PSP has a negative radial speed and is directly heading into the flow of outbound particles. On the post-perihelion arc, both PSP and β-meteoroids have positive radial speeds (PSP is "catching up" to the β-meteoroids) and the vector subtraction between their two velocity vectors leads to a smaller overall impact speed.

The two impact speed dependencies (bound vs. β-meteoroid) lead to substantially different expected impact rate trends. For circular impactors (or near circular from the dynamical model), the decrease in impact speed within the Kepler distance overcomes the increase in total density as radial distance decreases and causes a local depression in total impact rates at perihelion. For β-meteoroid rates, there is no such local minimum in impact speeds and the impact rates are expected to peak very near to perihelion. Additionally, the pre-perihelion arc is expected to have larger impact rates as the impact speeds are higher than at the equivalent heliocentric distance on the post-perihelion arc.

One surprising feature in the impact rate data is the apparent lack of a speed-dependent detection threshold. For dust grains on circular orbits, the number density as a function of size is typically assumed to be a power-law, such that $n(r, a) \propto r^{-1.3} a_{min}^{-3\alpha}$. Within sufficiently small size ranges, the interplanetary dust complex can be well-approximated by a power-law size distribution [e.g. Grün et al. 1985; Krivov et al. 2000]. Dust impact detection conventionally requires a specific quantity of impact charge to register a measurable signal. The impact charge produced by an impact is a product of the impactor mass and the speed to a large exponent, such that $Q \propto m v^{\gamma}$. For simplicity, we've assumed the impact charge scales linearly with impactor mass. The minimum detectable size then scales like $a_{min} \propto v^{-\gamma/3}$. Combining this with Eq. 13, the impact rate for a size dependent detection threshold is

$$R \propto r^{-1.3} v^{1+\alpha\gamma}. \qquad (14)$$

Faster particles produce more impact charge and are therefore more readily detected. Hence, a power-law size distribution leads to a speed-dependent minimum detectable size that varies with impact speed and modifies the detectable fluxes. For example, at higher speeds, smaller and more abundant particles are able to be detected and therefore the impact rates are enhanced. Likewise, at lower impact speeds, only larger and less abundant particles are detectable, and the impact rates are reduced.

The grey curve in Figure 7 shows expected speed-dependent impact rates for conservative size and speed exponents of $\alpha = 0.7, \gamma = 3$. In such a setup, even this conservative choice of exponents would predict a much larger decrease in impact rate near perihelion than the circular constant-size model rate shown in red in Figure 7. This is due to the diminished impactor speed, which significantly reduces the minimum detectable size and therefore the flux. However, a size-dependent circular orbit impact rate is not consistent with the impact rates observed by PSP. If FIELDS is registering impacts from dust grains on circular orbits, unless it is able to detect all impacts above ~20 km/s, impact rate profiles more similar to the grey curve would be expected.

As shown in Figure 7, the observed impact rates are more consistent with a dominantly β-meteoroid impactor population than for bound dust grains. The lack of speed-dependent detection threshold further reinforces this conclusion, where no dip is observed in total impact rate near perihelion as expected for slower, bound impactors. Additionally, as listed in Table 3, the directionality of impactors also provides a separate diagnostic on the dominant impactor source. While the impactor direction is more difficult to quantify than total impact rate, parallel efforts to characterize the impactor directionality show the directionality may also be consistent with β-meteoroids [Page et al., submitted].

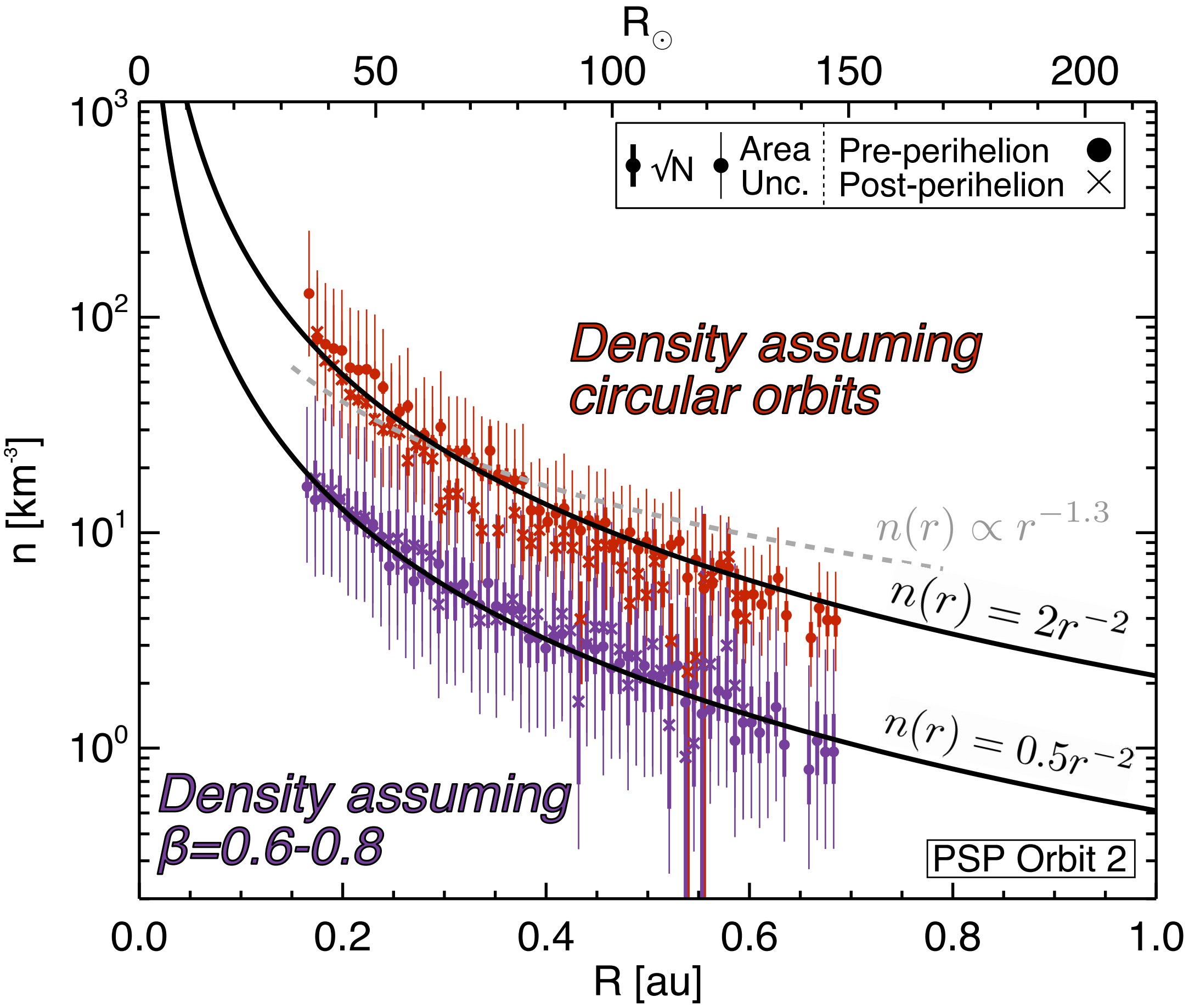

**Figure 8.** Derived dust radial profile from FIELDS-TDS data assuming either all impactors are on circular orbits (red) or β-meteoroids for PSP orbit 2. Thick error bars show Poisson $\sqrt{N}$ and thin error bars show the range of densities varying the expected area by a multiplicative factor of 2 on top of the Poisson errors.

From the impact rate profile, a radial density can be derived via $n(r) = R/Av_{imp}$, shown in Figure 8 for the TDS rates. Given the discrepancy between the directionality and radial trends, we derive densities assuming either all impactors are on circular orbits (red) or β-meteoroids (purple) for a smaller range of $\beta = 0.6 - 0.8$. Thick error bars show Poisson $\sqrt{N}$ and thin error bars show the range of densities varying the expected area by a multiplicative factor of 2 on top of the Poisson errors. Overlaid in grey is an example density curve that scales like $n(r) \propto r^{-1.3}$ [Leinert et al. 1978], which provides a poor fit to the densities derived assuming circular orbits. Both densities are found to be much more consistent with a $r^{-2}$ dependence, where the best fit of this scaling for each population is shown with the black lines.

Of the two densities, the β-meteoroid population is found to have a much higher fidelity fit than the circular derived density. For the circular derived densities, the pre-perihelion and post-perihelion arcs yield separate trends, where the post-perihelion derived densities are systematically lower than the pre-perihelion derived densities, and are inconsistent with an azimuthally symmetric zodiacal cloud. This would already be expected given the more precipitous decrease in the post-perihelion arc impact rates shown in Figure 7. Conversely, assuming all impactors are β-meteoroids gives a much higher fidelity density trend that is mostly devoid of systematic differences in derived densities between the pre- and post-perihelion arcs. This further suggests, from the radial trend, that β-meteoroids are the primary impacting flux detected by FIELDS.

### 3.2 ISΘIS Impact

The Integrated Science Investigation of the Sun (ISΘIS) instrument contains two energetic particle instruments [McComas et al. 2016]. One of these instruments, EPI-Lo, has 80 separate apertures, which view approximately half the sky at any given time. These 80 apertures each have a collimator foil directly exposed to space, which is recessed in an approximately cylindrical housing [Hill et al. 2017]. At 2019-093 16:45, aperature 31 (denoted L31 for Look direction 31) registered a permanent increase the noise floor. This noise increase is interpreted to be due to a dust particle penetrating the collimator foil and exposing the telescope to additional UV noise through the small puncture. The L31 telescope has a field-of-view (FOV) that is approximately looking in the orbital plane, centered ~12˚ away from the solar direction towards apex. The collimator has a full-width FOV of 56˚. The FOV has an approximately triangular area response function for the collimator foil, with a foil area of 36 $mm^2$. The bottom panel of Figure 9 shows the FOVs for all 80 aperatures, with L31 highlighted in white. The inset in the middle panel shows the approximate geometry of the collimator FOV and foil.

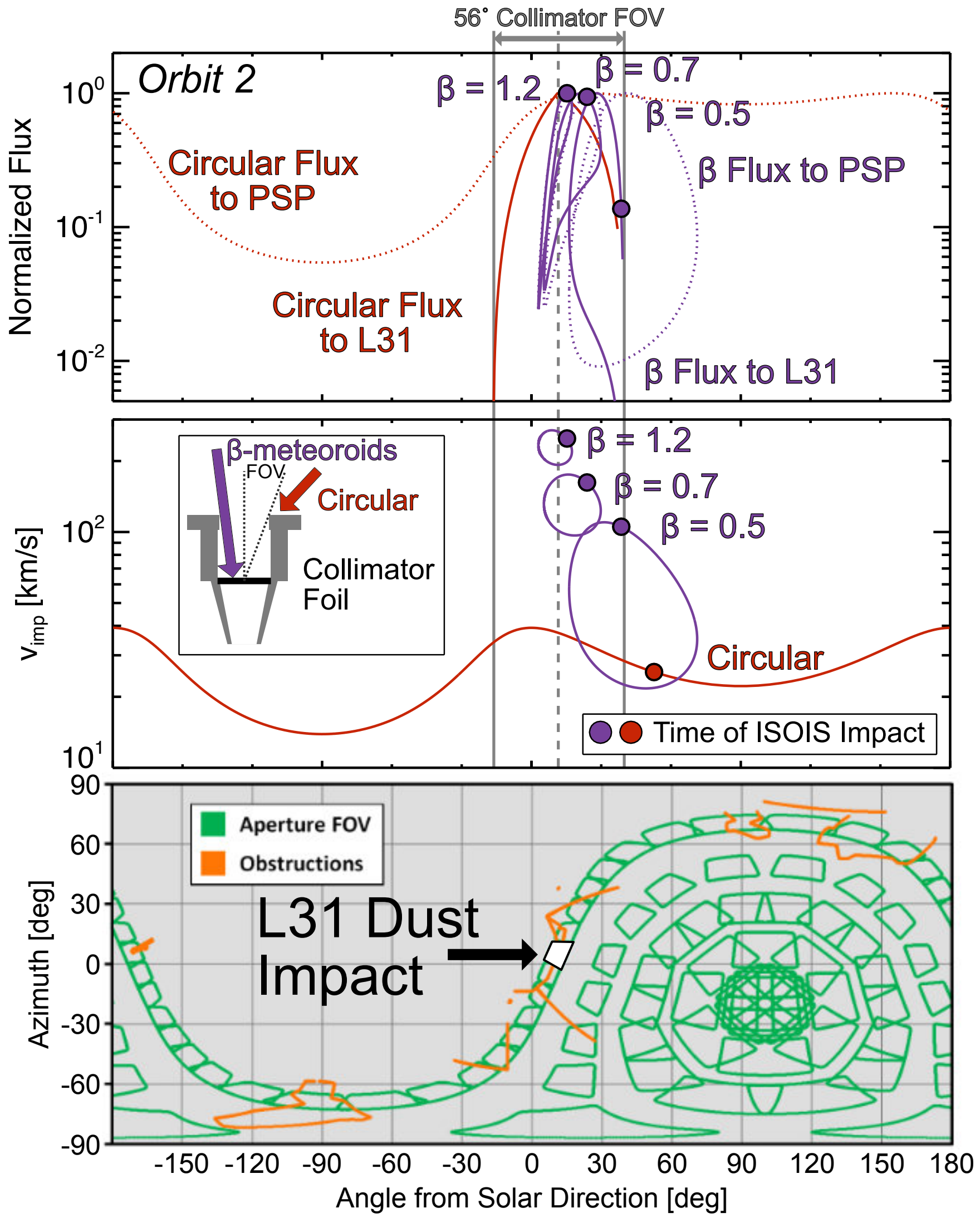

**Figure 9.** Top: normalized flux as a function of impact angle to PSP (dotted) and to the L31 collimator foil (solid) for circular (red) and β-meteoroids (purple). Middle: impact speed as a function of impact angle. Bottom: EPI-Lo aperture FOVs for all 80 look directions, with L31 highlighted.

The top panel of Figure 9 shows the normalized flux as a function of impact angle to PSP (dotted) and to the L31 collimator foil (solid) for circular (red) and β-meteoroids (purple). Colored dots indicate the expected value at the time of the ISOIS dust impact. The middle panel shows the impact speed as a function of impact angle. As shown in the top panel of Figure 9, the ISOIS impact is much more consistent with an impact from a β-meteoroid compared to a dust grain on a circular orbit. The flux of grains on circular, bound orbits at the time of impact is completely obscured by the collimator cylinder, as shown by the steep drop in the solid red curve in the top panel of Figure 9, where the impact occurs outside the circular FOV. The impact event to L31 well coincides with when L31 is expected to experience its peak flux from β-meteoroids over a range of $\beta = 0.5 - 1.2$. This time is also when the β-meteoroid impact speeds are near their peak values, at 100-250 km/s for $\beta = 0.5 - 1.2$ respectively.

If the IS⊙IS impact is due to a β-meteoroid from a population consistent with the FIELDS measurements, the total expected impacts can be estimated throughout PSP's orbit using Eq. 13. The top panel of Figure 10 shows the expected impact rates using the derived density fit for β-meteoroids from Figure 8. As shown in this figure, L31 is almost continually exposed to the flux of β-meteoroids throughout the orbit. This is due to it being pointed in a direction that happens to remain close to the β-meteoroid impact flux vector throughout the orbit. The bottom panel of Figure 10 shows the cumulative impacts expected to be incident on L31 throughout the orbit. For $\beta = 0.5 - 1.2$, we expect total impact counts of $0.2 - 1.5$ in orbit 2. Since orbit 1 and 2 are dynamically similar, we would have expected L31 to have a total cumulative impact count of $0.4 - 3.0$ over the first two orbits, consistent with a single impact to EPI-Lo's L31 collimator foil. Table 3 gives the expected total impacts per orbit to L31. While L31's collimator foil has most likely already been punctured by one dust impact, we would expect any additional puncturing impactors to this foil to produce a linear increase in UV noise. Therefore, L31, and all other collimator foils, can continue to serve as "dust detectors" after a single penetrating impact.

Since many of the EPI-Lo telescopes are pointed in similar directions, we would expect additional penetrating impacts to occur for other telescopes pointed near the β-meteoroid arrival direction during the PSP mission. A diagnostic for the relative importance of β-meteoroids in PSP's impactor environment will be to monitor and compare any additional dust impacts to the EPI-Lo collimator foils with the expectations in Table 3. Specifically, the timing and pointing orientation of exposed collimator foils upon impact provide key information on the impactor source for any future foil impacts.

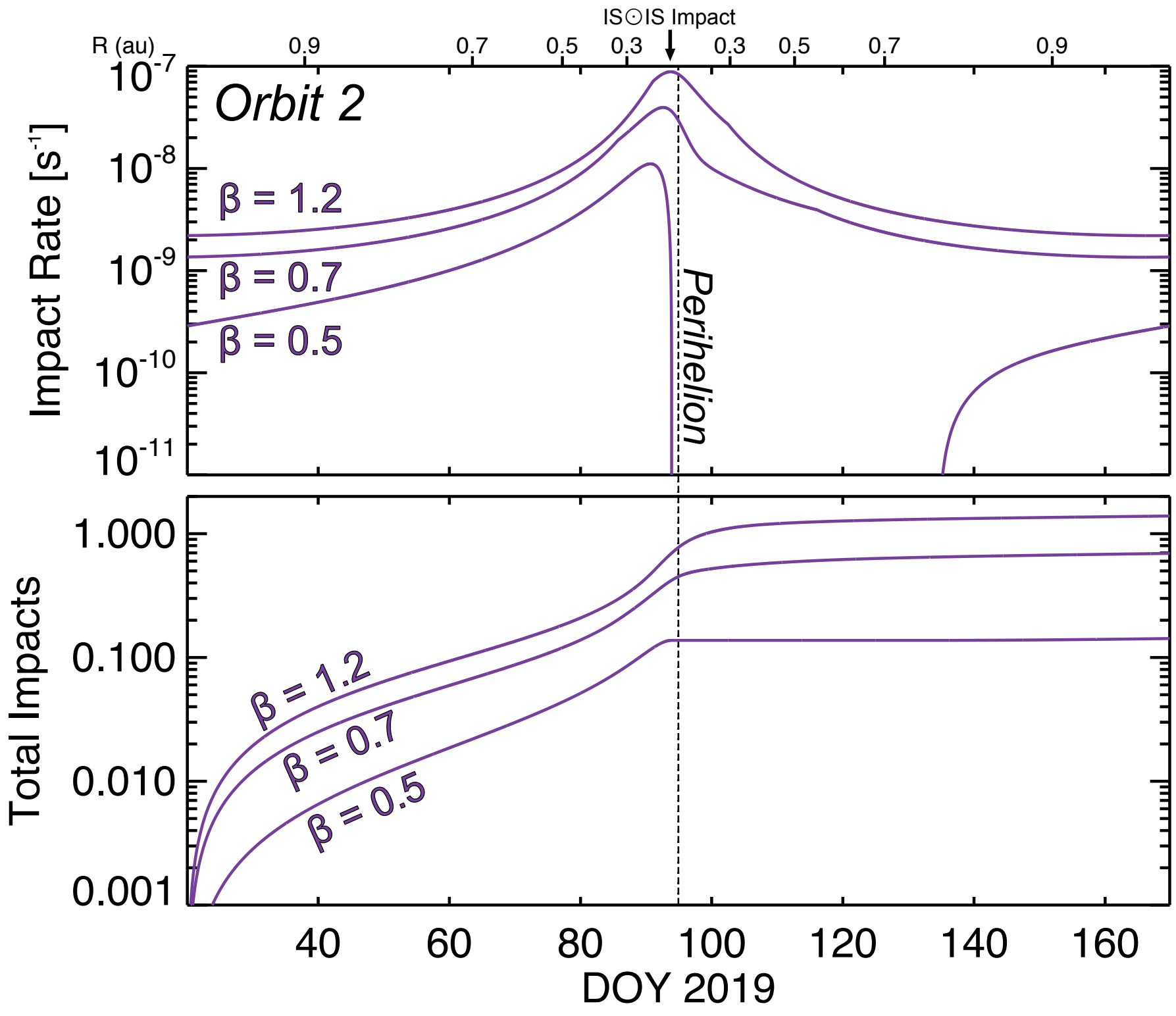

**Figure 10.** Modeled impact rates (top) and total cumulative impacts (bottom) to L31 throughout orbit 2 for $\beta = [0.5, 0.7, 1.2]$.

| Orbit | $\beta = 0.5$ ($n_0$ = 0.8) | $\beta = 0.7$ ($n_0$ = 0.5) | $\beta = 1.2$ ($n_0$ = 0.3) |
|---|---|---|---|
| **1-3** | 0.2 | 0.8 | 1.5 |
| **4-5** | 0.3 | 1.1 | 2.0 |
| **6-7** | 0.5 | 1.3 | 2.5 |
| **8-9** | 0.7 | 1.6 | 2.9 |
| **10-16** | 0.9 | 1.9 | 3.3 |
| **17-21** | 1.1 | 2.1 | 3.7 |
| **22-24** | 1.3 | 2.3 | 4.0 |

**Table 3.** Total expected impacts per orbit for L31 throughout the PSP mission for a selection of possible β values. $n_0$ density values at 1 au given in units of $km^{-3}$.

# 4. Discussion and Conclusions

Comparing the total impact rates observed by FIELDS to dust models, we find the impactor fluxes experienced by PSP are consistent with a population of unbound dust grains on hyperbolic trajectories, β-meteoroids. This conclusion is based on the strong match between total impact rates and model trends for β-meteoroids, along with the relatively poor match with expected impact rates from bound dust grains on circular orbits. The impact rates exhibit a significant asymmetry between the inbound and outbound arcs. Such a trend is predicted by the β-meteoroid model, where the inbound portion of the trajectory has larger impacts speeds and therefore receives a larger flux compared to the outbound leg. Bound impactors are expected to produce an approximately symmetric impact rate profile before and after perihelion, with a potentially small enhancement predicted for the post-perihelion arc from the dynamical model, opposite to the observed decrease in outbound impact rate. Additionally, the impact rate trend exhibits no appreciable dip near perihelion, which would be consistent with a speed-dependent threshold for lower-speed circular impacts. Furthermore, the derived densities are consistent with a density profile of $n(r) \propto r^{-2}$, which is expected for β-meteoroids, and are poorly fit with the bound density trend of $n(r) \propto r^{-1.3}$ [Leinert et al. 1978]. This trend suggests the impactor source is unlikely to be due to dust grains on bound orbits [Pokorny & Kuchner, 2019]. Finally, a degradation of one collimator foil on the ISΘIS/EPI-Lo instrument is consistent with an impact from a β-meteoroid and not from a dust grain on a circular, bound orbit.

A parallel analysis focusing on the inferred impactor directionality may also suggest the impact environment is consistent with β-meteoroids [Page et al. this issue]. Impactor directionality analysis in Page et al. [this issue] is based on the relative amount of impacts inferred to occur on PSP's ram vs. anti-ram hemispheres. In this study, we have made two key assumptions to compare the FIELDS impact rates with model trends in this work: 1) FIELDS is able to detect any impact on the spacecraft with equal efficiency for different materials and 2) the effective area available to detect impacts is the projected area of all spacecraft surfaces for a given dust impact flux direction. To determine the directionality information in the parallel study, additional assumptions must be made: 3) impacts closer to a given antenna will yield a larger charge signal and 4) the polarity of all impact signals are the same throughout the orbit [Page et al. this issue]. Better understanding the fidelity of these assumptions with subsequent orbits and increased statistics may reveal a more coherent story between the directionality and radial trend analyses.

While the data do not decisively arrive at one population as the dominant source given the discrepancy between radial and directionality trends, following the radial trend analyses outlined in this work, we investigate the implications if the impact rates are dominated by β-meteoroids. Assuming the observed dust impact rates are primarily due to impacts from β-meteoroids, the derived flux of β-meteoroids (product of density and radial velocity) at 1 au for $\beta = [0.5,1.2]$ is $0.3 - 0.7 \times 10^{-4}$ m$^{-2}$ s$^{-1}$. This is lower, but comparable to the flux of hyperbolic particles observed by Pioneers 8 & 9 over seven years of observations, with a flux of $2 \times 10^{-4}$ m$^{-2}$ s$^{-1}$ at 1 au [Berg and Grün 1973]. Discrepancies between these two fluxes may be due to the significantly different detection mechanisms used by Pioneers 8 & 9 and PSP. Additionally, dust impact measurements from the Helios mission have suggested a low mass density for β-meteoroids [Grün et al. 1980]. While the current set of PSP analyses cannot provide significant insight into the mass density of impactors, additional statistics may shed light on this difficult measurement.

Further following the assumption β-meteoroids dominate the observed PSP impact rates, the mass loss rate of the zodiacal cloud due to escaping β-meteoroids can be estimated as $\dot{M} = 4\pi r^2 n_\beta v_{\beta r} m_\beta$. Following the fitted trend, we can calculate this loss rate at 1 au. For a range of $\beta = [0.5,1.2]$ and mass density of $\rho = 2 - 3$ g/cc, $\dot{M} = 1 - 14$ tons/s. The total loss rate inside 1 au was previously estimated to be ~10 tons/s [Grün et al. 1985a], for which β-meteoroids were estimated to carry ~1-3 tons/s out of the solar system [Grün et al. 1985b], consistent with our findings. Our mass production estimates are also consistent with the upper end of additional modeling efforts that estimate β-meteoroid mass production of 0.03 – 5 tons/s [Mann & Czechowski, 2005]. Therefore, if the impact rates discussed in this study are dominated by β-meteoroids, their flux and total mass loss rates are consistent with existing estimates of β-meteoroid fluxes. Closer transits by the Sun may therefore reveal critical information about the source region for β-meteoroids. Additionally, these small dust grains may be a source for pickup ions near the Sun [Schwadron et al. 2000; Mann and Czechowski, 2005; Schwadron and Gloeckler, 2007] and their signature may be detectable by the plasma instrument (SWEAP) onboard PSP in future orbits.

The location of the assumed source region in this work of 5 $R_\odot$ is an approximation. β-meteoroid production due to collisions is occurring throughout the inner solar system. Approximately half of the zodiacal cloud collisions are expected to occur inside 0.18 au (39 $R_\odot$) [Zook and Berg 1975], and the zodiacal cloud fluxes near Earth are expected to reasonably be extrapolated down to 0.1 au (21.5 $R_\odot$) [e.g. Mann et al. 2004]. However, the location (and variability) of β-meteoroid production is still not fully understood and is model derived in lieu of spacecraft measurements deep in the inner solar system. If the PSP observed impact rates are dominated by β-meteoroids and if their production peaked farther out than PSP's perihelion distance of 0.17 au, we would expect to observe a significant flattening or depression in the impact rates near close approach, which is not shown in the data analyzed here. Therefore, we investigate the effects of modifying the assumed source location interior to PSP's perihelion distance.

We performed a sensitivity test and found the derived β-meteoroid density at 1 au increases by a factor of ~2.5 assuming all β-meteoroid production occurs at 0.16 au. However, the fidelity of the fit was decreased for this distance, as the density trends exhibited a systematic offset between the inbound and outbound derived densities, which is unphysical. Therefore, we conclude that for the orbit discussed in this work, the source region for β-meteoroids may have been significantly interior to 0.17 au. This does not preclude the possibility that the β-meteoroid production region can shift with variability in the zodiacal cloud. If there is sufficient variability in β-meteoroid production in the inner solar system, it is not possible to determine with a single orbit if the results derived in this work are indicative of the average behavior of the zodiacal dust cloud. Future PSP orbits will be able to address the variability and location of the production region for β-meteoroids. We note that future modeling efforts could also consider a distributed source region of β-meteoroids, compared to the single-distance simplified source region assumed here. Additionally, depending on the source region location for β-meteoroid production, this process may also be responsible for a chemical filtration of our solar system's zodiacal cloud as certain compositional features in the dust grains may be lost due to sublimation and sputtering [e.g. Mukai & Schwehm 1981, Mann et al. 2004] before their collisional breakup and subsequent ejection from the solar system as β-meteoroids.

If the impact rates are dominated by β-meteoroids as suggested in this work, this does not necessarily imply that the total zodiacal cloud number densities are dominated by β-meteoroids. Impact speeds for β-meteoroids to PSP are a factor of a few to an order of magnitude larger than those from circular, bound dust grains, hence their impact rates are significantly enhanced compared to grains on circular orbits. The relative densities of circular vs. β-meteoroids derived from fitting the PSP rates depend sensitively on the characteristic value of β. It may be the case that β-meteoroids make up a small portion of the total number density of the zodiacal cloud, but are much more readily detectable by PSP due to their significantly larger impact speeds. Sufficient statistics over many orbits may allow for a more accurate constraint on the range of β values and enable the relative densities to be constrained using PSP data.

**Acknowledgements**
The authors would like to thank the many Parker Solar Probe team members that made these observations possible. We also thank Eberhard Grün, Mihály Horányi, and Zoltan Sternovsky for helpful comments on the manuscript. We acknowledge NASA Contract NNN06AA01C. P.P. was supported by NASA Solar System Workings grant NNH14ZDA001N-SSW.